\documentstyle[prl,aps,epsfig]{revtex}
\begin{document}
\draft
\twocolumn[\hsize\textwidth\columnwidth\hsize\csname
@twocolumnfalse\endcsname 

\title{Tunneling of Bound Systems at Finite Energies: 
Complex Paths Through Potential Barriers.}
\author{G.F. Bonini$^a$, A.G. Cohen$^b$, C. Rebbi$^{b}$ and
  V.A. Rubakov$^{c}$}
\address{$^a$Institut f\"ur theoretische Physik, 
  University of Heidelberg, D-69120 Heidelberg, Germany}
\address{$^b$Department of Physics, Boston University,
  Boston, MA 02215, USA}
\address{$^c$Institute for Nuclear Research of the Russian
  Academy of Sciences, Moscow 117312, Russia}

\date{January 19, 1999}
\maketitle
\begin{abstract}
We adapt the semiclassical technique, as used in the context of
instanton transitions in quantum field theory, to the description of
tunneling transmissions at finite energies through potential barriers
by complex quantum mechanical systems. Even for systems initially in
their ground state, not generally describable in semiclassical terms,
the transmission probability has a semiclassical (exponential)
form. The calculation of the tunneling exponent uses analytic
continuation of degrees of freedom into a complex phase space as well
as analytic continuation of the classical equations of motion into the
complex time plane.  We test this semiclassical technique by comparing
its results with those of a computational investigation of the full
quantum mechanical system, finding excellent agreement.
\end{abstract} 
\vskip2pc]

1. Tunneling phenomena are inherent in numerous quantum systems,
from atoms to condensed matter to quantum field theory. Even in systems 
with a small parameter---coupling constant---a quantitative 
description of tunneling is possible only in a limited number of cases.
Perhaps the best known example is the WKB approximation 
familiar from one-dimensional wave mechanics; similar techniques, 
such as the ``most probable escape path'' and instanton 
methods~\cite{BBW,uses}, are used to study tunneling 
from the bottom of potential wells. In the latter cases the calculation 
of the tunneling probability may be reduced to the solution of classical 
equations of motion for {\it real} generalized coordinates 
in {\it imaginary} (``Euclidean'') time, supplemented by  the analysis of
small fluctuations about this classical Euclidean trajectory.
However, these methods often fail in describing tunneling of
systems with more than one degrees of freedom at finite energies.

It has been suggested recently~\cite{rt,rst}, in the context of 
instanton transitions in quantum field theory, that semiclassical
techniques  may be used for calculating the exponential
suppression factors in a class of processes where multi-dimensional
systems tunnel at finite energies. The proposal involves a double
analytic continuation: the degrees of freedom are continued 
into a complex phase space, and the equations of motion 
are solved along a contour in complex time.
The tunneling exponent is determined by an appropriate solution
of the calssical, albeit complexified, equations of motion.
Computation by numerical methods is then feasible even for systems with
a large number of degrees of freedom, as has already been demonstrated in a
field theoretic model~\cite{kut}. A problem with the formalism of
Refs.~\cite{rt,rst} is that its derivation from first principles is 
still lacking, although its plausibility has been supported by perturbative
calculations about an instanton \cite{tin,mul}.

The purpose of this paper is two fold. First, we adapt the technique of
Refs.~\cite{rt,rst} to tunneling of quantum mechanical
bound systems through high and wide potential barriers. As an example, 
we consider a system of two degrees of freedom with linear binding
force.  We find that if the bound system is initially 
in a highly excited state, the tunneling exponent is 
indeed calculable in a semiclassical way. This
result is hardly surprising,  as the initial state itself can 
be described in semiclassical terms. We formulate the complexified
classical boundary value problem relevant to the calculation of the
exponent in this case. 

Second, the real strength of this formalism is that it also enables one to
treat  barrier penetration when the bound system is initially in a 
{\it low lying} state, {\it e.g.}~the ground state. This is far from 
obvious, as this initial state {\it cannot} be described semiclassically.  
Nevertheless, we argue that in this case the tunneling exponent
can be obtained by an appropriate limiting procedure. The resulting
technique is less-well justified, so we have chosen to test it by direct
computation of the transmission probability in the full quantum theory.
We briefly describe the numerical methods involved, and present the
results of both the full quantum mechanical and semiclassical analyses.
We find good agreement between the two, confirming the validity of
the semiclassical approach. 
\vskip0.65pc
2. To be specific, let us consider a quantum mechanical system of
two particles of equal mass $m=1/2$ moving in one dimension. Let these
particles be bound by the harmonic potential 
$(\omega^2/8)(x_1 - x_2)^2$, and one of these particles be repelled 
from the origin by a positive semidefinite potential $V(x_1)$ that 
vanishes as $x_1 \to \pm \infty$ (we could of course allow $V$ to 
depend on $x_2$ as well, provided it couples to the internal 
degree of freedom).
We take this potential to have the form 
$V(x_1) = g^{-2}U(gx_1)$, where $g$ is a small constant. We set
$\hbar = 1$, so the classical limit corresponds to $g \to 0$.
In what follows we present the results of numerical calculations for
$\omega = 1/2$ and gaussian potential, $U(x) = \exp (-x^2/2)$,
although the treatment of other potentials would be similar.
In terms of the center-of-mass and relative coordinates,
$X=(x_1 + x_2)/2$ and $y = (x_1 - x_2)/2$, the Lagrangian reads
\begin{equation}
  \label{10*}
    L = \frac{1}{2} {\dot X}^2 + \frac{1}{2} {\dot
    y}^2 - \frac{1}{2} \omega^2 y^2 - 
    \frac{1}{g^2} U[g(X+y)]
\end{equation}
Far from the origin ($\vert X\vert  \to \infty$), 
the center-of-mass and internal degrees of freedom decouple 
and the system can be characterized by its center-of-mass momentum
$P$ and oscillator excitation number $n$, or, equivalently, by $n$ and the 
total energy $E=P^2/2 + \omega (n+ 1/2)$. We wish to calculate the
probablity $T_n(E)$ for transmission of the system through the barrier
$V$. Of particular interest is $T_0(E)$, the transmission probability of
this system  initially in its oscillator ground state.

It is convenient to introduce rescaled total energy and occupation number 
$\epsilon = g^2E$ and $\nu = g^2 n$. With our choice of 
$U$, the top of the barrier corresponds to a potential energy
$\epsilon =1$. For $\epsilon <1$ transmission is possible 
only via tunneling. For $\epsilon$ just above $1$ 
{\it classical} over-barrier transitions are possible for  very
special initial states. Indeed, there exists  an unstable, static classical 
solution with both particles stationary at the top of the barrier,
$x_1=x_2=0$,  so that $\epsilon = 1$.  If one perturbs this 
solution by giving an arbitrarily small, common positive 
velocity to both particles, they will move toward $X=\infty$. 
The reversed evolution takes the system to $X=-\infty$, 
with the classical oscillator characterized by a certain excitation energy
$\epsilon^{osc}_0 \equiv \omega \nu_0$ 
(and a certain phase of the classical oscillator). The combined evolution 
is the classical transition over the barrier from this particular asymptotic
state. By solving the (real time) classical equations of motion 
numerically, we found  that 
$\nu_0 \approx 0.9$  for $\omega = 1/2$.

The classical evolution of the system initially in the
classical oscillator {\it ground state}  ($x_1=x_2$) leads to the
excitation of the oscillator as it approaches the barrier. 
Classical transition over the barrier occurs in this case only if the
total energy exceeds some critical value. In our example we found 
numerically $\epsilon_{crit}= 1.8$. 

If $\epsilon$ and $\nu$ are such that classical transitions over 
the barrier are not possible, the system has to tunnel. We will shortly 
see that at $\epsilon$ and $\nu$ fixed, and $g \to 0$ (i.e., at large
total energy and initial occupation number, $E,n \propto g^{-2}$) the
transmission probability has the semiclassical form
\begin{equation}
  \label{13*}
    T_n(E)=C(\epsilon, \nu) e^{-\frac{1}{g^2} F(\epsilon, \nu)}
\end{equation}
The case of the initial oscillator  ground state is more subtle. 
In analogy to Refs.~\cite{rt,rst} we suggest that the transmission 
probability at $n=0$ has the form
\begin{equation}
  \label{13+}
    T_0(E)=C_0(\epsilon) e^{-\frac{1}{g^2} F_0(\epsilon)}
\end{equation}
and that the exponent is obtained by taking the limit
\begin{equation}
  \label{13**}
    F_0(\epsilon)={\rm lim}_{\nu \to 0} F(\epsilon,\nu) 
\end{equation}
One of the main purposes of this paper is to check this limiting
procedure by comparison with a fully quantum mechanical calculation.

\vskip0.65pc
3. To see that Eq.(\ref{13*}) is indeed valid, and to obtain the procedure
for calculating the exponent $F(\epsilon, \nu)$, let us consider the
transmission amplitude
$ A(X_f,y_f;P,n)=\langle X_f, y_f\vert \exp[- i H (t_f-t_i)] 
\vert P, n \rangle$, where $X_f$ ($> 0$) and 
$y_f$ are the coordinates at time $t_f$,
and we eventually take the limit $(t_f - t_i) \to \infty$.
This amplitude may be written as a convolution 
of the evolution operator in the coordinate basis
and the wave function of the initial state.
The former is given by the path integral 
$\langle X_f, y_f\vert \exp[- i H (t_f-t_i)] \vert X_i,
y_i\rangle = \int \![dX][dy]\, \exp{i S}$ where the integration
runs over paths satisfying $(X, y)(t_i) = (X_i, y_i)$, 
$(X, y)(t_f) = (X_f, y_f)$. For an initial state with  $P \propto
g^{-1}$, $n \propto g^{-2}$,  the initial wave function is 
semiclassical and has the exponential form.  In the case
of a harmonic binding potential,  this follows
from the integral representation in the coherent state formalism:
\[
 \langle X_i, y_i \vert P,n \rangle = \frac{e^{i PX_i}}{\sqrt{2 \pi}}
  \int\! {dz d{\bar z} \over 2\pi
       i }\, e^{-{\bar z} z} {{\bar z}^n\over \sqrt{n!}}
       e^{-\frac{1}{2} z^2 
       - \frac{1}{2} {\omega} y_i^2 +
       \sqrt{2\omega} z y_i}
\]
(One may replace ${\bar z}^n /\sqrt{n!}$ by
$\exp (n \log \bar{z}/\sqrt{n} + n/2)$ at large $n$.)
By introducing the rescaled integration variables $X \to gX$,
$y \to gy$, {\it etc.}, we observe that $ A(X_f,y_f;P,n)$ is given by an
integral of an exponential of the form
$\exp( - g^{-2} \Gamma)$ where 
$\Gamma$ depends only on the rescaled integration variables, $\nu$ and
$\epsilon$, and does depend explicitly on $g^2$. This allows for a
semiclassical analysis: we find stationary points of $\Gamma$ and
evaluate the integrals using a stationary phase approximation.
We outline the main steps in the derivation of the stationary point equations.

Variation of $\Gamma$ with respect to $X(t)$ and
$y(t)$ for $t_i < t < t_f$ leads to the conventional classical 
equations of motion. When classical transitions are forbidden,
there will be no real solutions satisfying the boundary
conditions. Nevertheless there will be solutions with complex 
values of the integration variables. When performing 
the analytic continuation we will, in general, encounter singularities.
To deal with this problem, we note that the {\it time} contour, originally 
the real axis, can be distorted into the complex plane, keeping the end points
$t_i$, $t_f$ fixed. This deformation of the time contour allows us to
avoid these singularities. Thus, our strategy is to search
for complex solutions of the classical equations of motion along a contour
ABCDE in the complex time plane, as shown in Fig.~1.

There are further stationary point equations coming from
variation of $\Gamma$ with respect to the integration variables at the
end point $t_i$. It is convenient to formulate these
equations along part B of the contour, where $t = iT/2 + t'$,
$t' = \mbox{real} \to - \infty$ (this is possible because the equations
of motion decouple in the asymptotic past). Instead of $\epsilon$ and
$\nu$ we introduce new real parameters $T$ and $\theta$; $T$ enters the
problem through the shape of the contour. The general complex solution at 
large negative $t'$ is $X(t') = X_0 + pt'$, 
$y(t') = u e^{-i \omega t'} + v e^{i \omega t'}$ where $X_0$, $p$,
$u$ and $v$ are complex parameters. The stationary point equations
at the initial time lead to the following boundary conditions:
(i) $X(t')$ is real ({\it i.e.} $p$ is real and $T$ may be chosen so
that $X_0$ is also real), 
(ii) the positive and negative frequency parts of $y(t')$ are related to
each other by $v = u^{*} e^{\theta}$. 

More bo\-und\-ary con\-diti\-ons appear when one evaluates 
the total transmission probability, {\it i.e.}~integrates 
$\vert A(X_f, y_f; P, n) \vert^2$ over $X_f$ and 
$y_f$, again in a gaussian approximation. These conditions involve the
final time and simply require that 
(iii) $X(t)$ and $y(t)$ are real on the DE part of the contour.

\begin{figure}[t]
\vspace{7mm}
\centerline{\epsfxsize=3 in \epsfysize=2 in \epsfbox{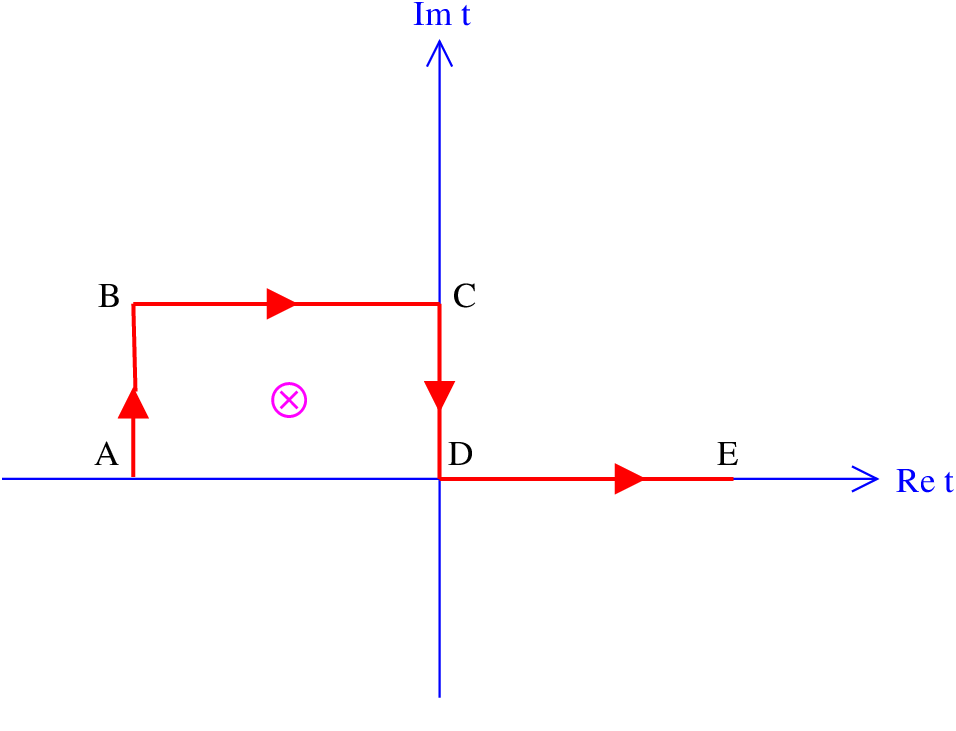}}
\noindent
Fig 1. {\it Complex time contour used to find the stationary point solutions.}
\end{figure}

At given $T$ and $\theta$ these three boundary conditions are sufficient
to specify the complex solution of the classical equations of motion
on the contour BCDE (up to time translations along the real axis).
Given this solution, the exponent for the
transmission probability (\ref{13*}) is the value of $2 \mbox{Re}\Gamma$ at
the stationary point. Explicitly, we find
\[
      F(\epsilon, \nu) = 2 \mbox{Im} S_0 - \epsilon  T - \nu\theta
\]
where 
\[
  S_0 = - \int_{BCDE}~dt~\left[ 
   \frac{1}{2} X \partial^2_t X
 + \frac{1}{2} y \partial^2_t y + \frac{\omega^2}{2} 
   y^2
 + U(X + y) \right]
\]
is the (rescaled) classical action for the complex solution
of the above boundary value problem. The total energy and 
excitation number are related to $T$ and $\theta$ by
\[
   \frac{\partial (2 \mbox{Im} S_0)}{\partial T} = \epsilon
   \;, \;\;\;
    \frac{\partial (2 \mbox{Im} S_0)}{\partial \theta} = \nu
\]
{\it i.e.} the pairs $(\epsilon, \nu)$ and $(T, \theta)$ are
Legendre-conjugate. 

We have solved the equations of motion numerically along the contour BCDE
subject to the boundary conditions (i)--(iii). In particular, we
have evaluated the limit (\ref{13**}). 
The result of this semiclassical calculation is shown in Fig.~3.

\vskip0.65pc
4. To check this semiclassical procedure, we have performed a numerical 
analysis of the full quantum system defined by
(\ref{10*}). This is conveniently done in a basis of
center-of-mass coordinate $X$  eigenstates  and oscillator excitation
number $n$. In this basis the state is represented by a multi-component
wave function $\psi_n(X) \equiv \langle X,  n \vert \Psi \rangle$,
and the time-independent Schr\"odinger equation reads
\begin{eqnarray}
\label{18*}
 -{\partial ^2 \psi_n(X) \over \partial X^2} + \left(n+{1 \over 2}\right)
 \omega \psi_n(X) &&
 \nonumber \\
+ \sum_{n'} V_{n n'}(X) \psi_{n'}(X) &=& E \psi_n(X)
\end{eqnarray}
where  $V_{n n'}(X)= \langle n \vert V(X+y) \vert n' \rangle$. 
Our choice of a gaussian potential $V$ enables us to calculate  
$V_{n n'}(X)$ by a numerical iteration procedure. Equation 
(\ref{18*}) is supplemented with the standard boundary conditions: 
(a) the incoming wave ($X \to - \infty$) is in a state of given 
center-of-mass momentum $P$ and excitation number $n$; 
(b) only outgoing waves exist at $X \to +\infty$.

To solve the system (\ref{18*}) numerically, we introduce a lattice with
equal spacing, $X_k = ka$, and discretize eq. (\ref{18*}) using
the Numerov--Cowling algorithm (which reduces the discretization error
to $O(a^6)$). We also truncate the system to a finite number of
oscillator modes $n \leq N_0$. In order to insure good accuracy 
of the solution, we have chosen the number of lattice 
sites $2N_X$ and the cutoff $N_0$ as large as $2N_X = 2\cdot 4096$,
$N_0 = 400$. 
This corresponds to over 3 million coupled complex equations. To deal with 
them, we take advantage of the special form of Eq.~(\ref{18*}). Indeed,
by inverting a set of $(N_0 + 1)\times (N_0 +1)$ matrices, which is
computationally feasible, Eq.~(\ref{18*}) can be recast in the form 
$\psi_n(X_k) = \sum_{n'} [L_k \psi_{n'}(X_{k-1}) +R_k \psi_{n'}(X_{k+1})]$.
The elimination of $\psi_n$ at definite $X_k$ leads to a system of
similar form for the remaining variables (with suitably redefined
$L$ and $R$), again after    $(N_0 + 1)\times (N_0 +1)$ matrix algebra
and matrix inversion. In this way we progressively eliminate variables at
intermediate values of $X_k$ and ultimately obtain a system that linearly
relates $\psi_n$ at the end points $X= - N_X a$ and  $X= + N_X a$.
With a discretized  version of the boundary conditions (a) and (b), this
final system is straightforward to solve. The transmission probability
is then determined by $\vert \psi_n \vert^2$ at the end point  $X= N_X a$.

We performed a series of checks of this numerical procedure
to insure that our calculations are sufficiently precise and that the
results are close to the continuum limit.

\vspace{5mm}
\begin{figure}[t]
\centerline{\epsfxsize=2.5 in \epsfbox{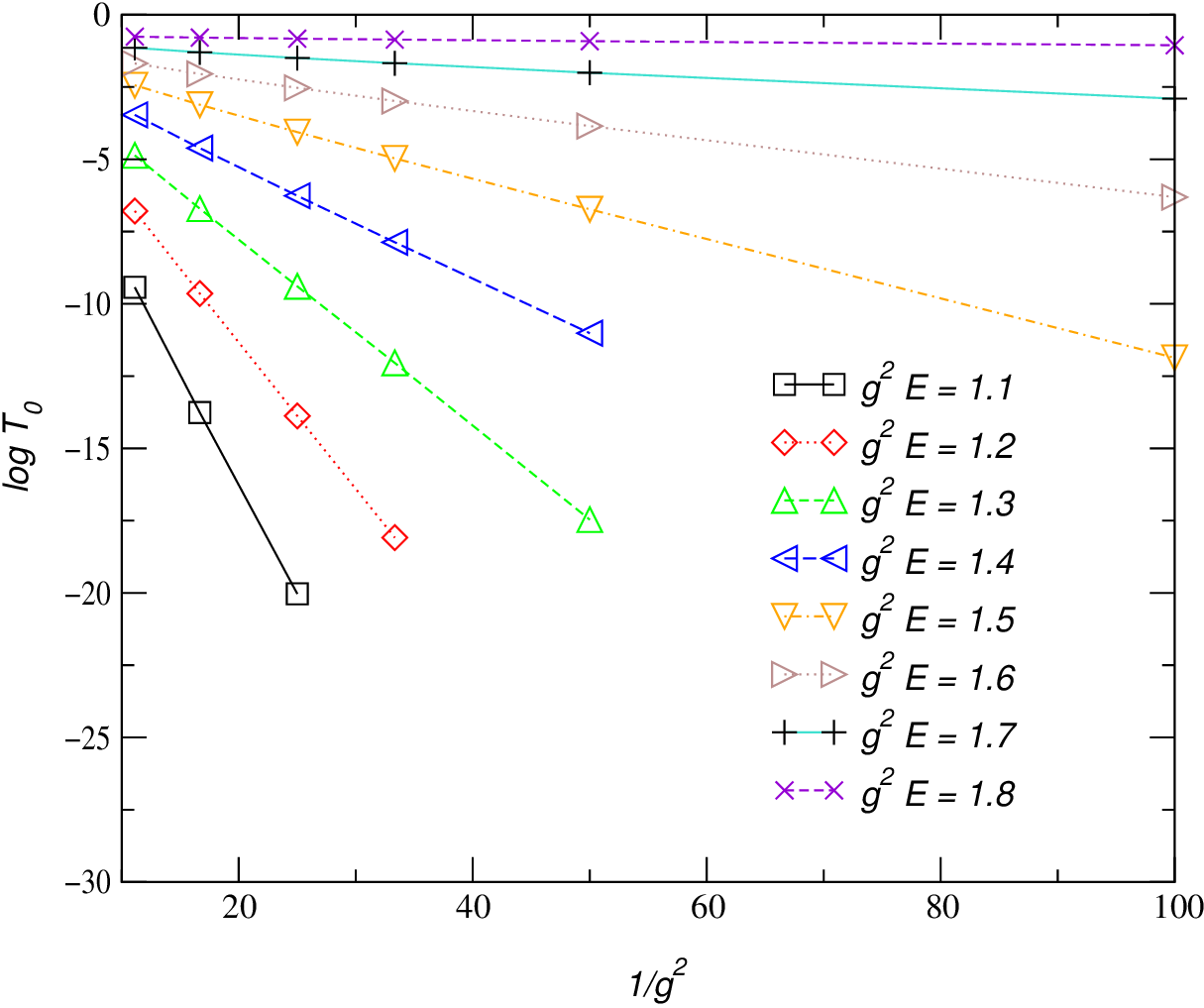}}
\noindent
Fig 2. {\it Logarithm of the transmission probability vs. $1/g^2$.}
\end{figure}

We present in Figs.~2 and 3 the results of the full quantum mechanical 
computation of the transmission probability for the system initially in 
its oscillator ground state.
The potential $V$ is gaussian, and $\omega = 1/2$. Figure 2 shows that the
transmission probability $T_0 (E)$ indeed has the functional form
(\ref{13+}): at fixed $\epsilon \equiv g^2E$, the logarithm of $T_0$ is
very well fit by a linear function of $g^{-2}$. We use this fit to
obtain the exponent $F_0(\epsilon)$. Both the full quantum mechanical
results for $F_0(\epsilon)$ and the semiclassical results (the latter 
obtained by implementing the limiting procedure (\ref{13**})) are shown 
in Fig. 3. Clearly, there is good agreement between the two. 
(The slight discontinuities in the  quantum mechanical
results are an artifact of the energy dependence of the 
$g^2$ range from which we can extract $F_0$. They
provide an indication of the errors due to higher order
effects.) We conclude that the validity of the semiclassical approach 
is confirmed by the direct quantum mechanical computation.

\vspace{7mm}
\begin{figure}[t]
\centerline{\epsfxsize=2.5 in \epsfbox{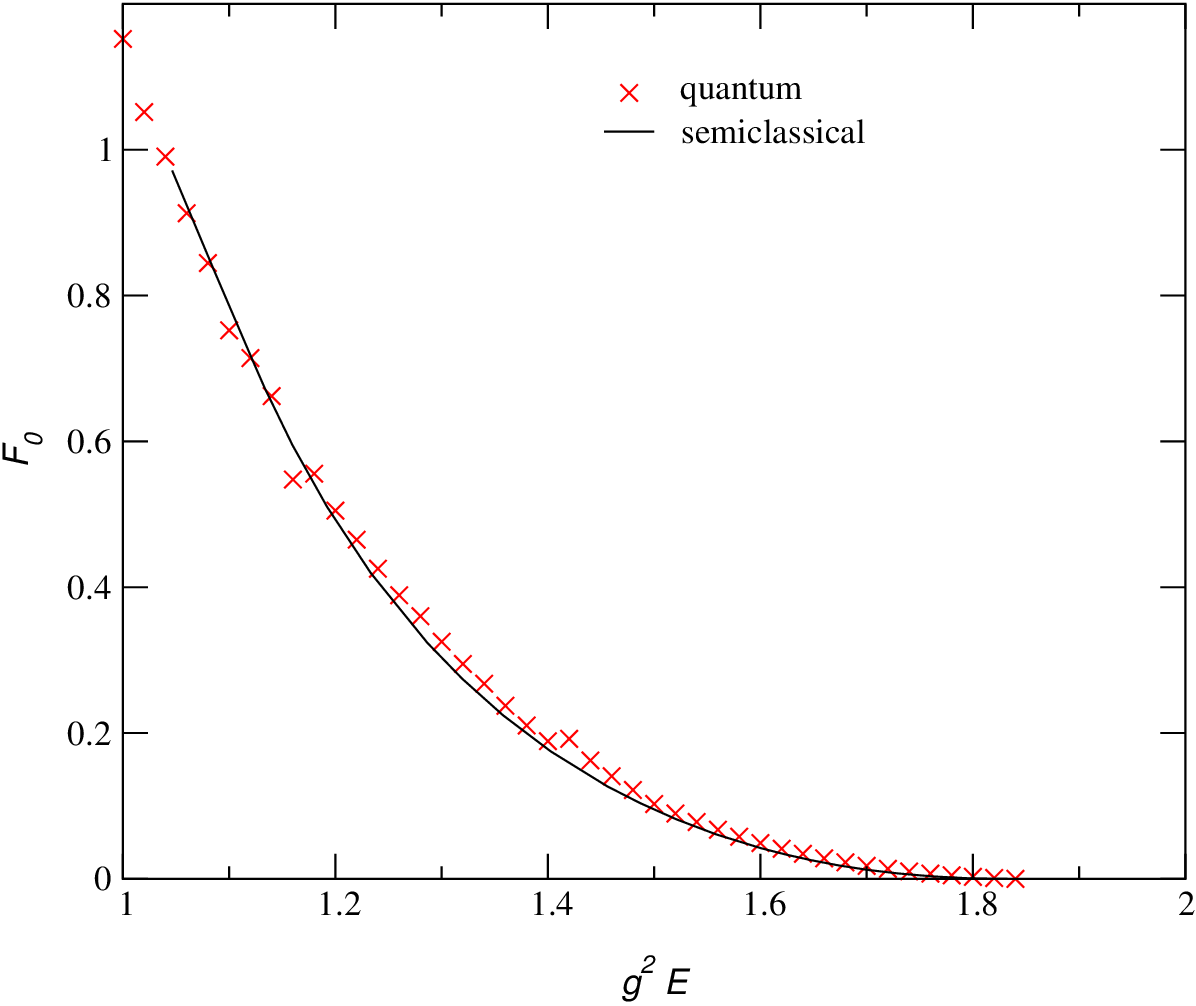}}
\noindent
Fig 3. {\it Quantum mechanical and semiclassical results.}
\end{figure}

\vskip0.65pc
5. Full quantum mechanical computations (analytic or numerical) of
barrier penetration probabilities are rarely possible. Even for our
simplified system, values of $g$ smaller than $0.1$ are difficult to study,
as one has to deal with very small transmission coefficients.
On the other hand, limitations of the semiclassical computations are 
far less severe. The generalization of the semiclassical approach to 
quantum-mechanical systems with harmonic binding of more than two 
particles in more than one space dimension is straightforward, and we also
expect that other binding potentials may be treated in a similar way
provided their semiclassical wave functions are known. Indeed, in all 
such cases the transmission amplitudes with highly excited initial states
will be given by (path) integrals of exponential functions, and the
tunneling exponents will be determined by appropriate stationary
points. The latter will be complex solutions to classical field 
equations on contours in complex time, with boundary conditions 
depending on the binding potential. A limit analogous to 
Eq.~(\ref{13**}) will then determine the tunneling exponent for 
incoming systems in low lying bound states.

The semiclassical calculability of pre-exponential factors is
less clear. While it is plausible that these factors are
given by functional determinants about complex classical solutions for
highly excited incoming states (finite $\nu$ in our model), 
we do not expect that  a limiting property similar to
Eq.~(\ref{13**}) will continue to hold for the pre-exponents. The  
calculation of such pre-exponential factors 
for low lying states remains an interesting open problem.

\vskip0.65pc
{\bf Ackowledgements.}  We are indebted to P.~Tinyakov for
helpful discussions. This research was supported
in part under DOE grant DE-FG02-91ER40676,
Russian Foundation for Basic Research grant 96-02-17449a and by the
U.S.~Civilian Research and Development Foundation for
Independent States of FSU (CRDF) award RP1-187.
Two of the authors (C.R. and V.R.) would like to thank Professor 
Miguel Virasoro for hopsitality at the Abdus Salam International Center 
for Theoretical Physics, where part of this work was carried out.

\end{document}